# FERROELECTRICS

## Title: Ferroelectricity in layered bismuth oxide down to one nanometer

**Authors**: Qianqian Yang[1#], Jingcong Hu[2#], Yue-Wen Fang[3,4#], Yueyang Jia[5], Rui Yang[5], Shiqing Deng[1], Yue Lu[2*], Oswaldo Dieguez[7], Longlong Fan[8], Dongxing Zheng[9], Xixiang Zhang[9], Yongqi Dong[10], Zhenlin Luo[10], Zhen Wang[8], Huanhua Wang[8], Manling Sui[2], Xianran Xing[11], Jun Chen[6], Jianjun Tian[1*], Linxing Zhang[1*]

**Affiliations:**

1. Beijing Advanced Innovation Center for Materials Genome Engineering, Institute for Advanced Materials and Technology, University of Science and Technology Beijing, 100083, China

2. Institute of Microstructure and Property of Advanced Materials, Faculty of Materials and Manufacturing, Beijing University of Technology, Beijing, 100124, China

3. Centro de Física de Materiales (CSIC-UPV/EHU), Manuel de Lardizabal Pasealekua 5, 20018 Donostia/San Sebastián, Spain

4. Fisika Aplikatua Saila, Gipuzkoako Ingeniaritza Eskola, University of the Basque Country (UPV/EHU), Europa Plaza 1, 20018 Donostia/San Sebastián, Spain

5. University of Michigan–Shanghai Jiao Tong University Joint Institute, Shanghai Jiao Tong University, Shanghai, China

6. Department of Physical Chemistry, University of Science and Technology Beijing, Beijing, 100083, China

7. Department of Materials Science and Engineering, The Iby and Aladar Fleischman Faculty of Engineering, The Raymond and Beverly Sackler Center for Computational Molecular and Materials Science, Tel Aviv University, Tel Aviv, Israel

8. Institute of High Energy Physics, University of Chinese Academy of Sciences, Chinese Academy of Sciences, Beijing 100049, P. R. China

9. Physical Science and Engineering Division, King Abdullah University of Science and Technology (KAUST), Thuwal 23955–6900, Saudi Arabia

10. National Synchrotron Radiation Laboratory, University of Science and Technology of China, Hefei 230026, China.

*11. Institute of Solid State Chemistry, University of Science and Technology Beijing, Beijing, 100083, China.*

*\*Correspondence to: linxingzhang@ustb.edu.cn, tianjianjun@mater.ustb.edu.cn, luyue@bjut.edu.cn*

*#These authors contributed equally to this work.*

**Abstract:** Atomic-scale ferroelectrics are of great interest for high-density electronics, particularly field-effect-transistors, low-power logic, and non-volatile memories. We devised a film with a layered structure of bismuth oxide that can stabilize the ferroelectric state down to one nanometer through samarium bondage. This film can be grown on a variety of substrates with a cost-effective chemical solution deposition. We observed a standard ferroelectric hysteresis loop down to ~1 nm thickness. The thin films ranging from 1 to 4.56 nm thickness possess a relatively large remanent polarization from 17 to 50 $\mu C \cdot cm^{-2}$. We verified the structure with first-principles calculations, which also pointed to the material being alone-pair-driven ferroelectric material. The structure design of the ultrathin ferroelectric films has great potential for the manufacturing of atomic-scale electronic devices.

**Main Text:**

Ultrathin ferroelectric film is a core material for the preparation of miniature and large-capacity non-volatile memories (*1,2*). The urgent demand for ultra-scaled devices has prompted the gradual exploration of atomic-scale ferroelectric thin films (*3*). In recent decades, we have witnessed that some traditional perovskite oxide systems (*4-8*), doped $HfO_x$ ferroelectric systems (*9,10*), and two-dimensional layered ferroelectric systems ($CuInP_2S_6$ (*11*), $α$-$In_2Se_3$ (*12,13*), etc.) keep the macroscopic ferroelectric gradually approaching the sub-nanometer size, but still far from the atomic-scale. The primary problem that hinders the continued research of nano-scale ferroelectric thin films is the critical size effect, where the huge depolarization field caused by the thickness reduction shields the ferroelectric effect, leading to the instability of ferroelectric phase(*1*). For example, as the thickness decreases to tens of nanometers or a few nanometers, ferroelectric thin films of classical perovskite with $ABO_3$ structure (i.e. $BaTiO_3$ (*7,8*)) will change from the ferroelectric phase to the paraelectric phase and lose the ferroelectricity. However, recent studies suggest that the size effect can be suppressed in some materials. Many films as thin as nanometers or atomic-scale (several unit cells) have been reported to still be ferroelectric, such as 3-unit-cell freestanding $BiFeO_3$ films with giant polarization due to the release of the tensile strain given by the substrate (*14*), 1-nm-thickness $Hf_{0.8}Zr_{0.2}O_2$ film with enhanced ferroelectricity due to imposed confinement strain (*9*), half-unit-cell (~3 angstroms) hafnium oxide with localized dipoles induced by flat phonon bands (*15*), and emergent ferroelectricity in sub-nanometer $ZrO_2$ films through reducing the dimension on

silicon(*16*). However, the ferroelectric properties of the ultrathin films for these works are only confirmed through cross-sectional high-angle annular dark-field scanning transmission electron microscopy (HAADF-STEM) images, piezoresponse force microscopy (PFM), theoretical calculations, or tunnel electroresistance hysteresis, rather than macroscopic ferroelectric hysteresis loop with polarization-electric field measurement, which can identify the ferroelectricity directly and is the prime determinant of the applications of ferroelectrics in electronic devices.

Layered bismuth oxides are one kind of classical ferroelectric materials with high Curie temperature ($T_c$) and great fatigue resistance (*17,18*). These systems of $Bi_2WO_6$ (*19*), $SrBi_2Ta_2O_9$ (*17*) and $Bi_4Ti_3O_{12}$ (*4*), etc., with a unique Aurivillius structure, are the famous layered bismuth oxides ferroelectrics (*20,21*), which are composed of an intergrowth of $(Bi_2O_2)^{2+}$ sheets and *n* perovskite-like blocks that contain a layer of octahedral *B* sites. A series of new Bi-based oxide ferroelectric films with a layered supercell structure have been reported recently, such as $Bi_2AlMnO_6$ (*22*), $BiMnO_3$, and $Bi_2NiMnO_6$ (*23*), which have a high degree of flexibility in its structure with the controlled bismuth layers. Some of them, for instance, show great multiferroic properties. Most of the above layered films exhibit in-plane ferroelectric properties, which limits their application in devices (*24*). The growth rate of the layered structure film in the horizontal direction is higher than that in the vertical direction, which is conducive to the preparation of smooth and continuous atomic-level films (*25,26*). The excellent insulating property and high tolerance to the vacancy of layered bismuth oxide are also beneficial to the ferroelectric measurement of monolayer unit-cells with low leakage (*27*). We have developed a type of layered bismuth oxide through Sm substitution (Bi$_{1.8}$Sm$_{0.2}$O$_3$ (BSO)). We grew this thin film as a single phase on a (0001) Al$_2$O$_3$ (AO) or (001) SrTiO$_3$ (STO) substrate by the sol-gel method. The film can maintain extremely strong out-of-plane ferroelectricity under 1 nm thickness, and demonstrate the macroscopic ferroelectric hysteresis loop that other systems cannot achieve below the thickness of one nanometer. This work develops a new generation of ferroelectric film, which is highly promising for creating miniaturized and high-quality electronic devices.

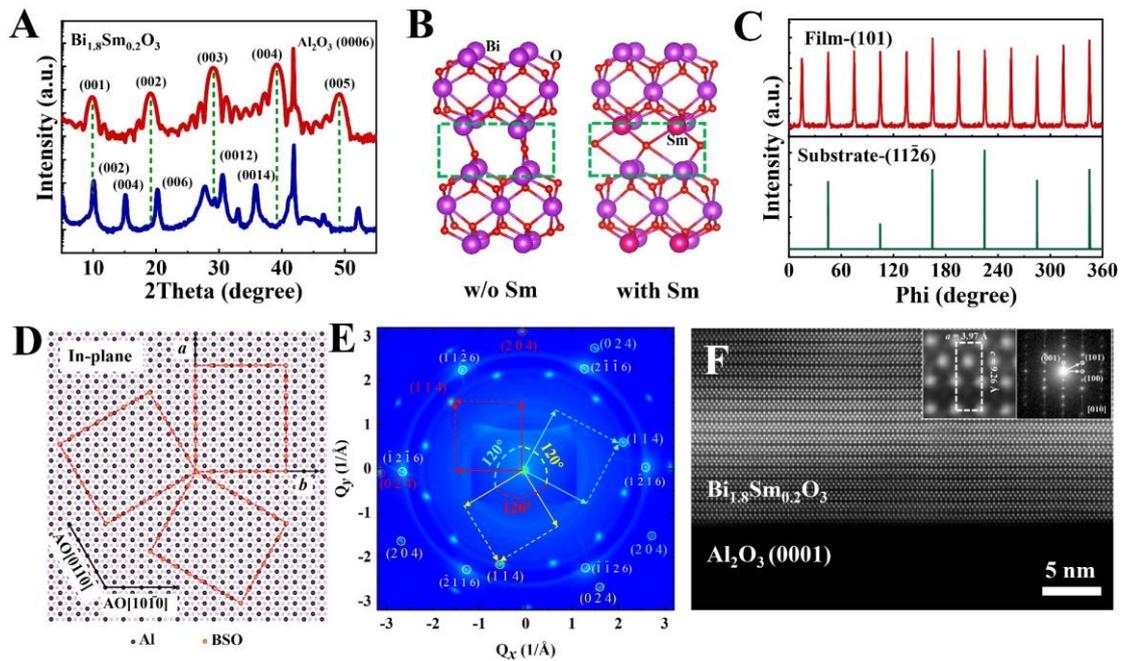

**Fig. 1 Crystal structure characterization of layered bismuth oxide thin films grown on (0001) Al$_2$O$_3$ substrates.** (**A**) Out-of-plane XRD diffraction of bismuth oxide films with (red) and without (blue) Sm substitution under the same conditions. (**B**) The final structures of Bi$_6$O$_9$ and Bi$_5$SmO$_9$ simulated in ab-initio molecular dynamics simulations at 300K. The green dashed rectangles are used to guide the eyes. (**C**) XRD Phi scanning reflections of (101) BSO film plane and (11$\bar{2}$6) Al$_2$O$_3$ substrate plane. (**D**) The matching diagram of the BSO film and the substrate (The red wire frame contains 6 (3) × 7 BSO

unit cells). (**E**) The diffractive RSMs images of film and substrate in ($Q_x$, $Q_y$) plane at $Q_z$ = 2.72Å$^{-1}$, where the substrate diffraction peak is marked in white, and the thin film diffraction peak is marked in red, blue and yellow. The three different colors correspond to the three domains with varied orientations respectively and show a rotational symmetry relationship of 120° between each other. (**F**) HAADF-STEM image of the heteroepitaxial BSO/AO, as viewed along the *c*-axis of the thin film. Inset in (**F**) are the zoom-in region of the film and the Fast Fourier Transform (FFT) of the film region.

**The design of layered structure**

The bismuth-based fluorite structure is a kind of classical material with variable and flexible structure (*28,29*). We designed a structure based on the fluorite structure through removing a whole layer of bismuth, leading to a layered structure with a framework of bismuth oxide. Density functional theory (DFT) calculations (Fig. S1) demonstrate that these layered structures with variable periods of different bismuth layers possess relatively high stability due to a low formation energy of -1.086, -1.135, -1.095 eV/atom for variable periods of 3, 5, 7 bismuth layers, respectively. We focus on the period of 3 bismuth layers, which can be stabilized in the film with a tetragonal-like (T-like) structure under the action of Sm substitution (Fig. 1A). If the oxygen is not removed in the fluorite structure with one in four Bi layers missing, the molecular formula is $Bi_6O_{16}$, which is not stable. Hence, we perform genetic-algorithm searches of the arrangements that lead to the lowest possible energy of each set of atoms in these cells of $Bi_6O_n$ (*n* = 6-12) with different oxygen loss (Fig. S2). Regarding formation energy by comparing the 69 known bismuth oxide compounds in the database, the $Bi_6O_9$ is strongly favored over the others, which also possesses a wide band gap and the most stable structure. In fact, the preparation of the film requires the addition of Sm element to stabilize the structure. We used energy dispersive spectrometer (EDS) to analyze the composition of the film and (001) STO substrate, respectively, and the chemical composition ratio we obtained was Bi:Sm:O = 1.8:0.2:3 (Figs. S3-S4). In addition, the valence analysis of X-Ray Photoelectron Spectroscopy (XPS) confirms the results of EDS, which was similar to the atomic ratio of the precursor (Fig. S5, Table S1).

**General structures**

We grew BSO films on an inexpensive single crystal (0001) AO substrate by chemical solution method of sol-gel spin coating (*30*). The X-ray diffraction (XRD) patterns showed only diffraction peaks from the directions of (0006) AO substrate and (001) films of BSO with an out-of-plane lattice parameter of about 9.137 Å (Fig. 1A). This observation suggests that the films are epitaxially grown, which is also confirmed by the Phi scans and reciprocal space mappings (RSMs) discussed below. To further confirm that the presence of Sm can stabilize the formation of this phase, we investigated varying the content of Sm from 5% to 15% (Fig. S6). The film without the Sm substitution consists of a main phase of $Bi_6O_7$ and some $Bi_2O_3$ phases, highlighting the importance of Sm in stabilizing the ferroelectric structure. Ab-initio molecular dynamics simulation results show that Sm has stronger oxygen binding ability than Bi (Fig. 1B, Fig. S7). When the thickness is down to one unit cell, the ferroelectric phase can be maintained by the addition of Sm (Fig. S8). Moreover, the introduction of Sm reduces the structure formation energy by 0.41 eV/atom. We used XRD phi scanning, wide-range RSMs and FFT to reveal the in-plane lattice matching relationship between BSO films and hexagonal AO substrates. The XRD phi scans are collected along the reflection of the (101) BSO film and the (11$\bar{2}$6) substrate (Fig. 1C). The phi scan of the (101) planes of the film show 12 diffraction peaks, of which 6 diffraction peaks correspond to the (11$\bar{2}$6) diffraction peaks of the hexagonal AO substrate. This features a four-axis symmetric structure of films with three different orientations on the six-axis symmetric substrate, confirming an in-plane epitaxial relationship with the substrate (Fig. 1D). We provide a wide range of RSMs of the sample in the ($Q_x$, $Q_y$) and ($Q_x$, $Q_z$) planes, respectively (Fig. 1E, Fig. S9). The relationship between thin film and substrate diffraction peaks shown by RSMs is consistent with Phi scanning results. RSMs about the (105) plane of the film has been carried out (Fig. S9C), indicating a T-like phase structure with in-plane lattice parameter of 3.94 Å and out-of-plane lattice parameter of 9.24 Å.

The direct evidence for the layered structure in the present film is provided by the spherical aberration-corrected HAADF-STEM images (Fig. 1F, Fig. S10A). We can observe flat and high-quality films. The bright spots correspond to Bi (Z = 83, where Z is the atomic number)/Sm (Z = 62) columns, owing to the $Z^2$-dependent image contrast (*31*). Because Sm is randomly distributed in the position of Bi, it is difficult to distinguish the exactly occupation site of them (Figs. S11-S14). The film is arranged in a regular layered arrangement with a period of three Bi-O layers, being similar to the fluorite structure along the *c*-axis direction. A large spacing between each two groups of three Bi-O layers, indicating a different layered structure of bismuth oxide than previously observed. The out-of-plane and in-plane lattice parameters are 9.26 Å and 3.97 Å, which corresponds to the results of XRD and RSMs. The corresponding FFT pattern (inset of Fig. 1F), identified as the [010] zone axis, which is consistent with the simulated BSO electron diffraction pattern (Fig. S10B). This confirms that the film has a T-like phase structure, which is the same as the RSMs scan result. The continuity and periodicity of the phase along the *a* and *c* axes can be clearly seen in the FFT. In addition, we analyzed the epitaxial relationship between the BSO film and the AO substrate by adjusting the focus and contrast (Figs. S15-S16).

Interestingly, the growth of this layered BSO film is extremely flexible and has little dependence on the types of substrates. Under the same conditions, the high-quality BSO film can also be successfully obtained on other lattice-mismatched substrates with different crystal structures, such as (001) STO and Au/SiO$_2$/Si substrate (Figs. S17-S18). We confirmed the epitaxial orientation matching relationship of BSO/AO and BSO/STO along [001]/[0001] and [001]/[001] (Figs. S19-S20). We obtained the full width at half maxima (FWHM) of $\omega$-scanning for the film and the substrate at a specific crystal plane to prove crystallinity (Fig. S21). The FWHM of (004) BSO film and (0006) AO substrate are approximately 0.0122° and 0.0123°, respectively. The FWHM of (004) film grown on the STO substrate and (002) STO substrate are approximately 0.0291° and 0.0303°, respectively. The FWHM of the films grown on these two substrates are almost the same as those of the respective substrates, indicating that BSO films grown by chemical solution method generally have a high crystalline quality whether it is grown on a single crystal (0001) AO substrate or (001) STO substrate. This would be ascribed to the design idea of layered structure based on the spontaneous arrangement in the in-plane direction of crystal atoms.

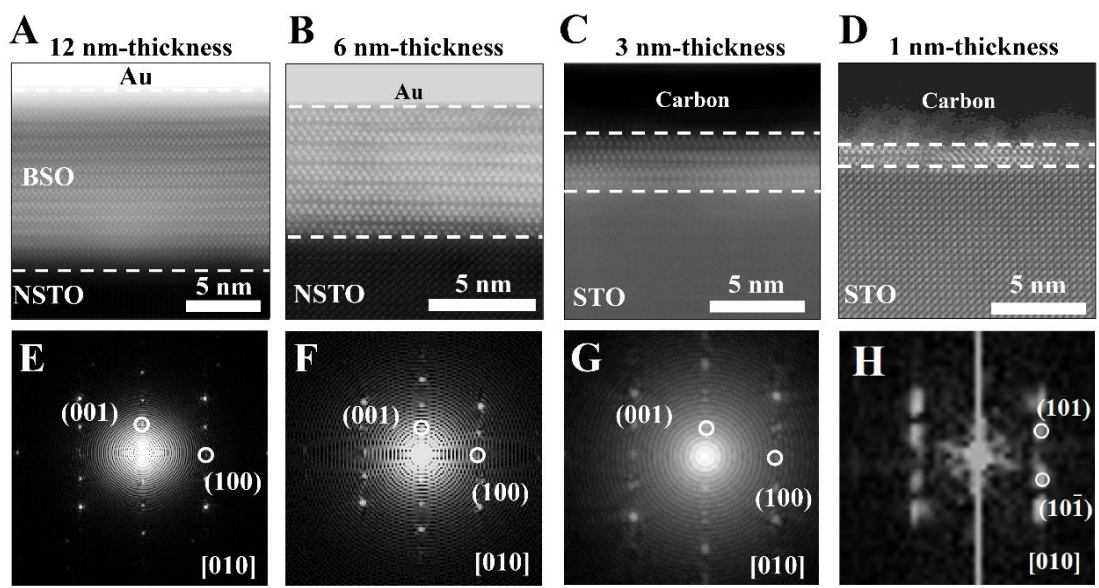

**Fig. 2 Characterization of ultrathin BSO films.** (**A**)-(**D**) The HAADF-STEM images of the BSO film grown on the STO-based (STO or Nb:STO (NSTO)) substrate have thicknesses of 12 nm, 6 nm, 3 nm, and 1 nm, respectively. (**E**)-(**H**) are the FFT patterns corresponding to the (**A**)-(**D**), respectively.

**STEM analysis for ultrathin BSO film**

We demonstrated that the film still has high crystalline quality and high flatness in the ultrathin state. After finding the proper growth conditions, we prepared a batch of single-phase BSO films with thicknesses of 12 nm, 6 nm, 3 nm, and 1 nm on the STO-based substrates. The thickness of the sample can be controlled by the concentration of the precursor, with a linearly relationship (Figs. S22-S23). Their HAADF-STEM images and corresponding FFT patterns are shown in Fig. 2. First, we directly observed that all BSO films have high flatness without showing the characteristics of "ripples", and the BSO films are continuously arranged in Bi atomic layers (Fig. 2A-D, Fig. S24).

Remarkably, the structure of the BSO film does not collapse when approaching 1-unit-cell thickness (1 nm-thickness), but remains layered and the chemical ratio is still maintained (Fig. 2E-H, Figs. S25-S26).

Second, the FFT patterns of all BSO films show bright and regularly arranged diffraction spots (the positions of diffraction spots are the same as Fig. 1F), which show that they all grow in the form of single crystals. X-ray reflectivity (XRR) and XRD fitting results also confirmed the high crystal quality and smooth interface of BSO films (Fig. S23, Table S2). As the thickness of the film decreases, the diffraction peaks move to low angles, indicating the gradually increasing lattice parameter *c*, which is similar to the results of TEM (Fig. S23, Table S3). We could attribute this observation to the compressive strain caused by the surface effect acting on the film (*32*). In addition, we also found a strain concentration region at the Bi layer at the BSO/AO (STO) interface, while the strain of other Bi layers extending towards the *c* axis is much weaker than that at the interface (Fig. S27). The successful preparation of high-quality BSO films with thickness of 1 nm is very likely due to the fact that the two Bi layers at the interface bear most of the strain from the substrate to maintain the normal growth of the films, which also lays a foundation for the measurement of the macroscopic ultrathin ferroelectric properties.

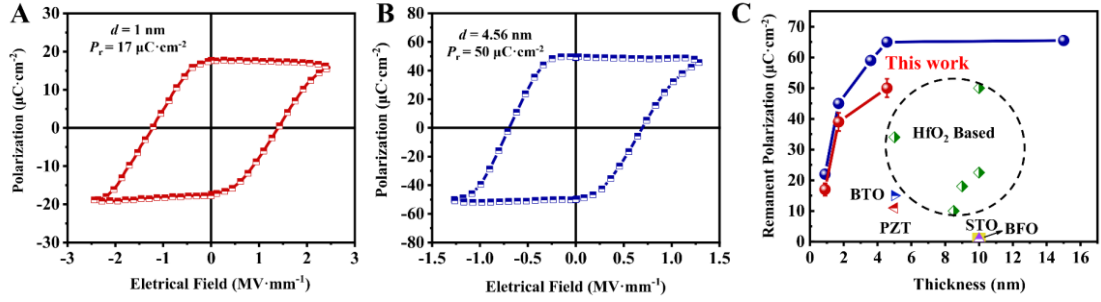

**Fig. 3 Characterization of macro ferroelectricity.** (**A** and **B**) The polarization hysteresis loops of the films with thicknesses of 1 nm (**A**) and 4.56 nm (**B**) obtained by the positive up negative down (PUND) mode measurement under an electric field with a frequency of 1 kHz. (**C**) The relationship between the thickness of the BSO film and the remanent polarization and the comparison with other ferroelectric systems (such as traditional perovskite systems and doped hafnium oxide systems). The blue solid symbols and the red solid symbols in the plot represent the dynamic hysteresis mode (DHM) conventional measurements and PUND measurements, respectively. In PUND mode, the contribution of non-ferroelectric switching can be subtracted to obtain true ferroelectric polarization (Figs. S28-S29). The error bars denote standard deviations.

**Ferroelectric hysteresis loop**

Ferroelectric hysteresis loops of ultrathin films are extremely difficult to measure. Influenced by size effect and film quality, the suppression of surface charges and the existence of leakage current causes the film to lose its original ferroelectricity. We obtained a high-quality ferroelectric hysteresis loop in the BSO film with a single-unit-cell thickness of 1 nm, with a remanent polarization as large as 17 μC·cm$^{-2}$ (Fig. 3A, Fig. S30A). Making this sort of observation in a traditional perovskite system, a doped hafnium oxide system or a two-dimensional ferroelectric system has been challenging at with ultrathin films down to 1 nm thickness. The ferroelectricity of the BiFeO$_3$ film with freestanding 3-unit-cell thickness (*14*), Hf$_{0.5}$Zr$_{0.5}$O$_2$ with a thickness of 1 nm (*9*) and CuInP$_2$S$_6$ film with a thickness of 4 nm (*11*) are all indirectly characterized by domain switching observed by applying a bias voltage. The macroscopic ferroelectric hysteresis loop is direct evidence to confirm whether the single-unit-cell thick film has ferroelectricity, which suggests that BSO ferroelectric thin films have great potential for applications in nano-electronic devices. The remanent polarization of the BSO film at 1 nm is relatively higher than that of other ultrathin ferroelectric films reported, such as BaTiO$_3$ (BTO, 5 nm, 15 μC·cm$^{-2}$) (*8*), BiFeO$_3$ (BFO, 10 nm, 1 μC·cm$^{-2}$) (*33*), PbZr$_{0.2}$Ti$_{0.8}$O$_3$ (PZT, 5 nm, 11 μC·cm$^{-2}$) (*34*), SrTiO$_3$ (STO, 10 nm, 1 μC·cm$^{-2}$) (*35*). The remanent polarization of the BSO film with a thickness of 4.56 nm increases to 50 μC·cm$^{-2}$ (Fig. 3B, Fig. S30B), which is the highest value among ultrathin ferroelectric film below 5 nm, compared to the hafnium oxide based ferroelectric films reported (5 nm, 34 μC·cm$^{-2}$) (*10,36-38*) (Fig. 3C). By using the semi-empirical method (*39*), as a displacement type ferroelectric, we calculated the polarization by the displacement. The atomic displacement of Bi is about 0.22 angstrom through STEM image analysis (Fig. S32), which is close to the theoretical prediction of 0.20 angstrom. We calculated the spontaneous polarization of the film to be about 49.8-53.4 μC·cm$^{-2}$. We obtained more hysteresis loops for the devices, indicating the ferroelectric repeatability of the films (Fig. S30). In addition, by adjusting the parameters in PUND mode, we obtained hysteresis loops under different conditions to obtain more ferroelectric information (Fig. S31).

As with traditional perovskite systems, the ferroelectricity of the BSO film is also affected by the thickness. The remanent polarization decreases with the reduced thickness (Fig. 3C). The size effect is probably due to the existence of a certain "dead layer" in the film (*40,41*). The "dead layer" usually has a low dielectric constant, resulting in a decrease in the actual voltage assigned to the thinner film and a smaller remanent polarization. Nevertheless, the polarization value obtained at present films is much higher than the conventional value. There are some reasons for the high polarization of present films at 1 nm. First, Sm can preserve the original structure of ferroelectric phase at one nanometer thick (Fig. S7). The layered structure and the high tolerance to vacancy are also beneficial to maintain ferroelectric structural stability at low dimensions. The stability of the ferroelectric phase will overcome the structural instability caused by the increase in surface energy and defects due to the size reduction. At the same time, the ferroelectric phase stability can also overcome the increase of depolarization field to maintain high polarization. Second, the polarization can enhance with increasing axial ratio when the thickness reduces. The displacement polarization also can increase from surface to interface. These results should be ascribed to strain effects and charge transfer (*42*) (Fig. S33), which will overcome the polarization attenuation caused by size effect.

The excellent fatigue and retention characteristics of atomic-scale BSO films and the virtually non-imprinting phenomenon at high temperatures confirm the great potential of BSO films for a wide range of device applications (Fig. S34). $T_c$ is an important parameter for ferroelectric materials. We have tried to explore the potential phase transition of this novel ferroelectric compound by XRD, dielectric, ferroelectric and theoretical calculation, which indicate that the $T_c$ of ultra-thin BSO films should be

493 K (Fig. S35).

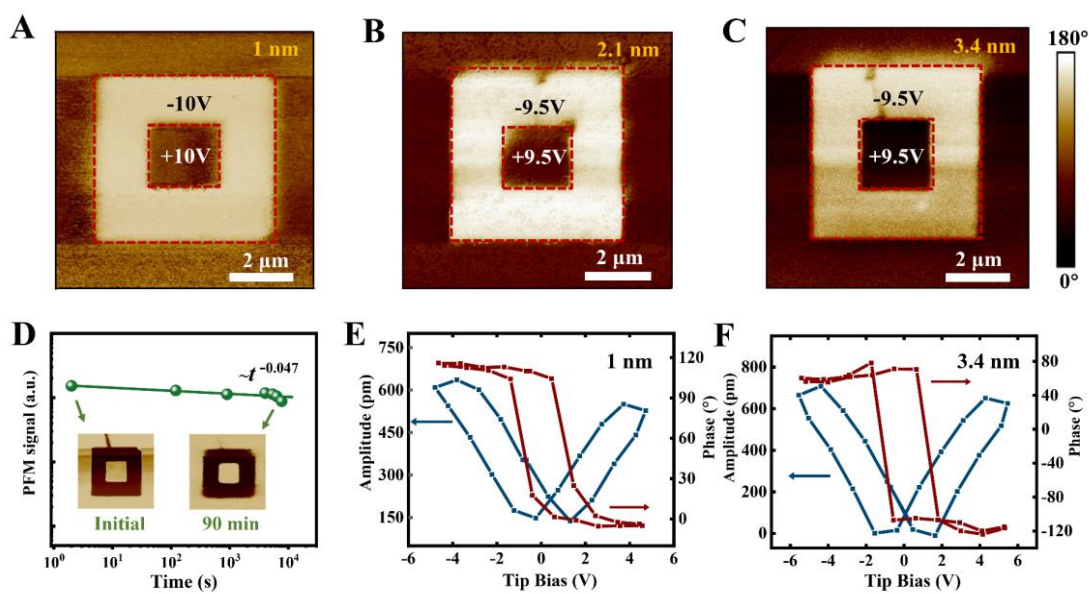

**Fig. 4 PFM of BSO thin films.** (**A-C**) The out-of-plane phase image after box-in-box writing with a tip bias in the BSO thin films with thicknesses of 1 nm (**A**), 2.1 nm (**B**) and 3.4 nm (**C**), respectively, by piezoresponse force microscopy. The entire detection area is 8 μm × 8 μm. (**D**) PFM signal as a function of delay time measured at room temperature after switching in a 1 nm-thick BSO film. The initial and 90 min later phase images are shown in insets. (**E-F**) Local PFM measurement results for films with thicknesses of 1 nm (**E**) and 3.4 nm (**F**), respectively, including both amplitude and phase hysteresis loop.

**PFM characterization**

We carried out the PFM measurement to investigate the ferroelectric switching properties of the present films. The PFM phase images after a box-in-box writing with a tip bias of positive and negative demonstrate well-defined regions of phase contrast, which corresponds to remanent polarization states, for BSO films with thicknesses of 1 nm, 2.1 nm, and 3.4 nm (Fig. 4A-C). This indicates the polarization states can be rewritten, highlighting switchable polarization for the ultrathin films especially down to one-unit cell, which could be also confirmed by the amplitude images (Fig. S36). Notably, unpoled areas at the edge show the similar phase contrast as positively poled areas, demonstrating that original BSO films exhibit spontaneous polarizations. In the respective square area, we selected multiple straight lines across the area where all voltages are applied to observe the phase lag of the overall area (Fig. S37). The results show that the phase lags of the films with thicknesses of 1 nm, 2.1 nm, and 3.4 nm are 70°-80°, 150°-160° and 160°-180°, respectively (Table S4). Then we randomly choose a position and measure the local amplitude and phase hysteresis loop (Fig. 4E-F, Fig. S38). Under various voltages, all the BSO films with different thicknesses show neat butterfly curves.

Surface topology inspection before and after PFM measurement and analysis of local PFM data after changing AC amplitudes and DC measurement frequency are analyzed to prove the ferroelectric nature in the BSO thin films (Figs. S39-S41). In addition, BSO film with a thickness of 1 nm shows excellent retention properties (Fig. S42, Fig. 4D). The PFM signal-Time diagram predicts that the polarization of the film can be maintained for several days or even longer after fitting. The solid line is the simulation curve of signal decay, which is in line with the power-law decay, $P(t) \propto t^{-\alpha}$, where $\alpha$ is decay exponent. That of the 1 nm thick BSO film is 0.047, which is far better than the reported conventional 12 unit-cells-thick ferroelectric $BaTiO_3$ film ($\alpha = 0.14$) (*35*). The long-term ferroelectric retention indicates that BSO films with a thickness of 1 nm have excellent ferroelectric stability, which is different from that of non-ferroelectric films. All piezoelectric responses of the above films prove the ferroelectric switching. The topography obtained by PFM proves that the BSO film prepared by chemical method has high quality. The BSO film with a thickness of 1 nm grows uniformly, and the root-mean-square roughness ($R_q$) is 0.105 nm (Fig. S37A, Table S4).

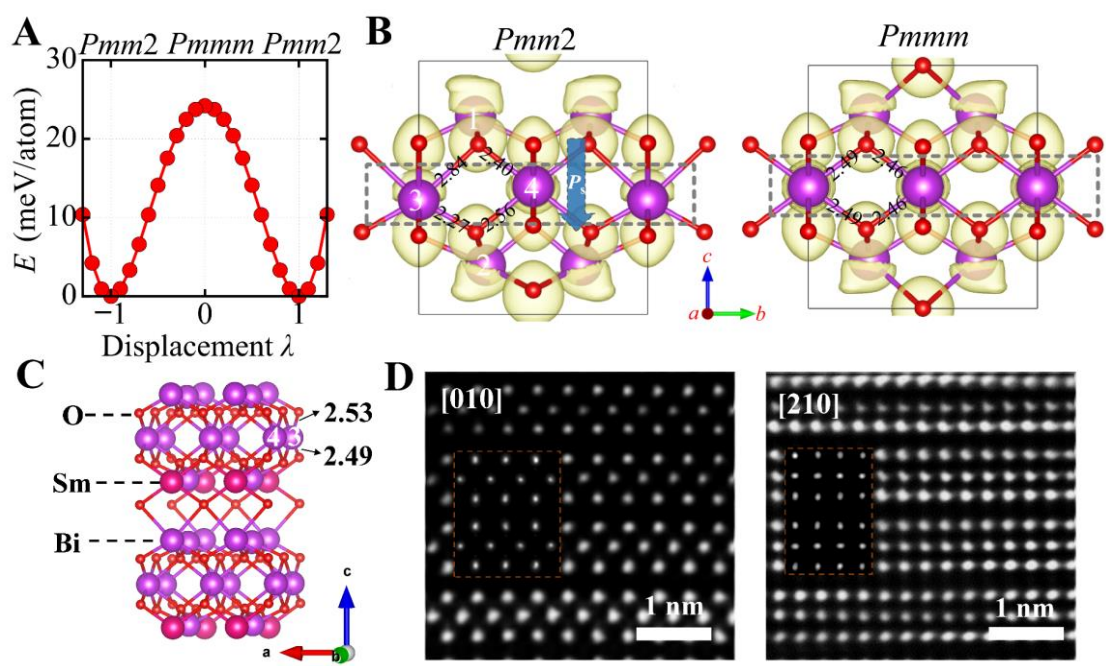

**Fig. 5 The BSO structure confirmed by DFT and HAADF-STEM.** (**A**) Total energy of $Bi_6O_9$ as a function of the normalized polar displacements $\lambda$ of $Bi_6O_9$ under the epitaxial strain of 3.97 Å. (**B**) The electron localization function (ELF) of ferroelectric and paraelectric phases of $Bi_6O_9$. The isovalue is set to 0.53. The dramatic change of ELF is highlighted by the dashed rectangles. The floating numbers label the bond lengths of Bi-O, and the integer numbers label the four inequivalent Bi atoms. The arrows point in the direction of spontaneous polarization. (**C**) The ferroelectric structure of Sm substituted $Bi_6O_9$ predicted by DFT. (**D**) The HAADF-STEM images of [010] and [210] directions as a comparison with the simulated HAADF images basing on the structure model (Orange-reddish-brown dotted box).

**Theoretical calculation for the structure and ferroelectricity**

To further understand the atomic structure, we performed a high-throughput DFT

crystal structure prediction within the structure space of $Bi_6O_9$ combining the genetic-algorithm implemented in USPEX, with information from the HAADF-STEM structures. In this section, unless otherwise specifically noted, all the structures are constrained to match the in-plane lattice constants of substrate with T-like lattice constant of 3.97 Å. Fig. S43A shows the formation energy profile of 120 predicted structures with the chemical formula $Bi_6O_9$. Several low-lying structures are predicted to be non-centrosymmetric (see the crystal structures in Fig. S43B-D). In particular, we found the structure in Fig. S43D with space group *Pmm*2 to open a band gap of around 1.2 eV by hybrid DFT calculations, indicating excellent insulating performance. The specific structural details of *Pmm*2 $Bi_6O_9$ can be found in Table S6, and the detailed densities of states calculated by different exchange functional with/without inclusion of spin-orbit coupling (SOC) are available in Fig. S44 and Fig. S45. For materials to be regarded as a ferroelectric material, it first needs to be an insulator. In addition, it is better to have a continuous and smooth transition between the non-centrosymmetric state and the centrosymmetric phase for the displacive ferroelectrics. By considering the group-subgroup relations (*43*), a phase with the space group of *Pmmm* is found to be the corresponding centrosymmetric structure of the *Pmm*2 $Bi_6O_9$. The specific structure of *Pmmm* $Bi_6O_9$ is provided in Table S6. Fig. 5A displays the free energy variation of $Bi_6O_9$ as a function of normalized polar displacements λ between the centrosymmetric and non-centrosymmetric structures, exhibiting a typical double-well energy landscape available in ferroelectric materials. The energy per atom monotonically decreases from the centrosymmetric (paraelectric) structure to the non-centrosymmetric (ferroelectric) structure, indicating continuous and spontaneous paraelectric-ferroelectric transitions under the critical temperatures. Strikingly, the energy difference between the centrosymmetric structure and the non-centrosymmetric structure is about 24 meV/atom (comparable to BTO (*44*)), which implies the relatively high stability of the polar phase of $Bi_6O_9$. Combining the double-well energy landscape with the insulating properties in electronic structure calculations, we have proven that the $Bi_6O_9$ is a new oxide ferroelectric. Therefore, the modern polarization theory based on the Berry phase method can be applied to $Bi_6O_9$. The spontaneous polarization is calculated to be around 30 μC·cm$^{-2}$, consistent with the polarizations measured in our thin film samples (17~50 μC·cm$^{-2}$).

In the Bi-based ferroelectric materials, the lone pair electrons of Bi play a large role in the spatial inversion symmetry breaking by exploiting so-called lone-pair stereochemical activity (*45*). To study the ferroelectric mechanism of $Bi_6O_9$, the electron localization function (ELF), which is a powerful tool to identify the role of lone pair electrons in driving the nonpolar-polar transitions (*46,47*) is used to reveal real-space visualization of the lone pair electrons. The ELF of *Pmm*2 and *Pmmm* $Bi_6O_9$ are comparatively shown in Fig. 5B. There are four symmetry-inequivalent positions of Bi atoms in the ferroelectric phase and therefore are marked as "1~4" in Fig.5B. In the transition from the paraelectric *Pmmm* phase to the ferroelectric *Pmm*2 phase, the ELF of Bi-3 and Bi-4 in the $O_8$ hexahedron cages transform from the spherically symmetric distribution to the lobe-like asymmetric distribution, which implies that the lone pair electrons of Bi-3 and Bi-4 are the driving force of the ferroelectricity by breaking the symmetry. We found the ELF of Bi-1 and Bi-2 are also slightly changed during phase transition. However, their ELFs are oppositely distributed along the +*c* and -*c* directions, which causes the dipoles to cancel each other out mostly and only results in a very small contribution to the ferroelectricity. The roles of lone pair electrons of Bi can also be understood by analyzing the element projected DOS and orbital projected DOS of Bi in Fig. S45 because the lone pair electrons are believed to stabilize the ferroelectricity through *sp* hybridizations (*48,49*). We find the *s* states of Bi-3 and Bi-4 are much stronger than those of Bi-1 and Bi-2 in the range from the energy of -2 eV to the Fermi level, which implies that the Bi-3 and Bi-4 have stronger hybridization with the *p* states than Bi-1 and Bi-2. This is consistent with the ELF analysis and further confirms the significant role of lone pair electrons of Bi-3 and Bi-4 in driving the ferroelectric transition. Next, to simulate the Sm substituted $Bi_6O_9$, we considered all the possible substituted configurations in which one Sm substitutes for one Bi within the unit cell of *Pmm*2 $Bi_6O_9$, the detailed comparison among of all the possible structures can be found in Fig. S46. The best structure match corresponds to a lowest-lying configuration and its crystal structure is shown in Fig. 5C. By analyzing the interatomic distances between Bi and O for Bi-3 and Bi-4 in Fig. 5C, Bi-3 and Bi-4 are found to be deviated from the center of $O_8$ hexahedron cage, indicating polarization remains in the Sm substituted $Bi_6O_9$. In addition, the density of states calculated by hybrid functional HSE06 in Fig. S47 show the Sm substituted $Bi_6O_9$ opens a band gap of 2 eV. Our calculations thus suggest that the ferroelectricity is preserved in the presence of Sm substituted in the $Bi_6O_9$. To compare the predicted crystal structure with the actual samples, we compare the crystal structure of Sm substituted $Bi_6O_9$ with the images and found it well reproduces the atomic structures on the [010] and [210] directions revealed by HAADF-STEM images, which confirms the accuracy of the predicted structure (Fig. S48). In order to confirm the crystal structure of BSO film with thickness of 1nm, we compared the predicted structure of $Bi_5SmO_9$ in Integrated Differential Phase Contrast (iDPC)-STEM (Fig. S49). At the same time, Fig. 5D shows the HAADF-STEM images simulated along the directions of BSO film [010] and [210], which are highly consistent and used to support the accuracy of the predicted structure. The simulation model is shown in Fig. S50. In addition, we observed that matching between BSO film and STO substrate is achieved by Bi-O and Ti-O layer of substrate. Therefore, a possibility of charge transfer between $Bi^{3+}$ and $Ti^{4+}$ at the interface exists, which would be a factor for the enhanced displacement polarization at the interface as described above.

**Conclusions**

We designed a ferroelectric with layered structure of bismuth oxide, and prepared the BSO thin film with good crystallinity through the sol-gel method, which can be grown on a variety of substrates. The film can still achieve macro-polarization at a thickness of 1 nm at room temperature, with high remanent polarization of 17 $\mu C \cdot cm^{-2}$. We confirmed the ferroelectric properties with measurements of writing domain and local butterfly curves in PFM. We obtained the structure of the BSO film by the DFT calculations, and confirmed that it is a different type of room-temperature ferroelectric film than previously observed. This provides a promising route for future research of ferroelectric materials, and these ultrathin ferroelectric films are highly suitable for future nano-electronic devices.

**Acknowledgments:** Author Contributions: L.X.Z. and J.J.T. conceived the idea of the work; L.X.Z. and Q.Q.Y. designed the research; Q.Q.Y., L.X.Z., J.C.H., Y.Y. J. and R. Y. fabricated the films and performed the initial tests. Y.-W.F. and O.D. performed and interpreted the theoretical calculations. J.C.H., Y.L., M.L.S., S.Q.D., D.X.Z. and X.X.Z. performed and analyzed the STEM experiments. Z.W. and H.H.W provided the fitting of XRR and XRD. Y.Q.D and Z.L.L performed and analyzed the XRD results. L.L.F., J.C. and X.R.X. conducted the ferroelectric properties measurements. Q.Q.Y. and L.X.Z. wrote the manuscript with contributions from others. All authors discussed the results and commented on the manuscript. L.X.Z., Y. L. and J.J.T. guided the projects.

This work was supported by the National Key Research and Development Program of China (2018YFA0703700, 2017YFE0119700 and 2020YFA0406202), the National Natural Science Foundation of China (21801013, 51774034 , 51961135107, 62104140, 12175235, 22090042, 12074016, 11704041 and 12274009), the Fundamental Research Funds for the Central Universities (FRF-IDRY-19-007 and FRF-TP-19-055A2Z), the National Program for Support of Top-notch Young Professionals, the Young Elite Scientists Sponsorship Program by CAST (2019-2021QNRC), and Lingang Laboratory Open Research Fund (grant LG-QS-202202-11). Use of the Beijing Synchrotron Radiation Facility (1W1A beamlines, China) of the Chinese Academy of Sciences is acknowledged. Y.-W.F. acknowledges the support of Masaki Azuma's group during his stay at the Tokyo Institute of Technology. Y.L. acknowledges the support of the Beijing Innovation Team Building Program (Grant No. IDHT20190503), the Beijing Natural Science Foundation (Z210016), the Research and Development Project from the Shanxi-Zheda Institute of Advanced Materials and Chemical Engineering (2022SX-TD001) and the General Program of Science and Technology Development Project of Beijing Municipal Education Commission (KM202110005003).

All data are presented in the main text and supplementary materials. There are no any competing interests including patents related to the work.

**Supplementary Materials:**

Materials and Methods

Figures S1-S50

Table S1-S6